\documentclass[12pt,a4paper]{article}
\usepackage{amsfonts}
\usepackage {graphicx}
\usepackage{amsmath}
\usepackage{amssymb}
\usepackage{amsthm}

\begin{document}

\font\ams=msam10 scaled\magstep1
\def\trid{{\ams \char'110 }}
\def\triu{{\ams \char'116 }}
\font\amss=msbm10  
\def\R{{\amss \char'122 }}
\def\C{{\amss \char'103 }}
\def\N{{\amss \char'116 }}
\def\Z{{\amss \char'132 }}
\def\Q{{\amss \char'121 }}

\newtheorem{theorem}{Theorem}[section]  

\newtheorem{corollary}[theorem]{Corollary} 

\title{Evaluating deterministic policies in two-player iterated games}

\date{}

\author{Rui Dil\~ao and  Jo\~ao Graciano}

\maketitle

\centerline{Non-Linear Dynamics Group, Instituto Superior T\'ecnico} 
\centerline{Department of Physics, Av. Rovisco Pais, 1049-001 Lisboa, Portugal}
\centerline{rui@sd.ist.utl.pt}
\bigskip

\begin{abstract}
We construct a statistical ensemble of games, where in each  independent subensemble we have two players playing the same  game.
 We derive the mean payoffs per move of the representative players of the game, and we evaluate all the deterministic policies with  finite memory. 
In particular,
we show that if one of the players has a generalized tit-for-tat policy,
the mean payoff per move of both players is the same, forcing the equalization of the mean payoffs per move of both players. In the case of symmetric, non-cooperative and dilemmatic  games, we show that generalized tit-for-tat policies together with the condition of not being the first to defect, leads to the highest  mean payoffs per move for the players. 
\end{abstract}


\section{Introduction}\label{s1}

Game theory has been formalized  by Neumann and Morgenstern in 1944, [1].  
Their objective was to introduce into the language of economic theory some mathematical tools for the quantitative analysis of the behaviour of economic agents without a central authority. One of the Neumann and Morgenstern arguments in favour of  the usefulness of a theory of games is based on the
intrisic limited knowledge about the facts which economists deal with. This argument has also been applied to
the description of some physical systems and to  evolutionary theories in biology. 

In the context of evolutionary biology and in order to analyse the logic of animal conflict, Maynard-Smith [2] introduced a game theory approach to describe some of the evolutionary features of organisms.  In the framework of sociology,  Axelrod [3] gave several examples where the game theoretical framework is useful. In economics, there is today a vast literature on the applicability of the game theoretical approach to economic decision [4], [5] and [6]. More recently, the same type of formalism has been applied to quantum mechanics [7].

In a game,  a policy is a rule of decision for each player, and policies  can be deterministic, depending of
the previous choices of one or of both players, or can be stochastic.
In a two-player  non-cooperative game with a finite number of choices or
pure strategies, 
both players know the payoffs,   make their choices independently of each
other, and know the past history of their choices.  It is also
assumed that
each player  maximizes its payoff after a finite or an infinite number of choices or moves.  

An important problem is game theory is to determine which policies perform better than others. In this context,
Axelrod, [3], proposed the following problem: "Under what conditions will cooperation emerge in a world of egoists without central authority?". To help to answer this question a computer tournament has been settled to decide which policy would perform better in an iterated Prisoner's Dilemma game, introduced by Dresher, Flood and Tucker [8]. The tournament has been won by the tit-for-tat (TFT) policy, submitted by Rapoport, [3, pp. 31]. The TFT policy consists in a simple rule that says that one's actual move is equal to what the other player did in the previous move.

In fact, several approaches have been developed in order to decide which policies perform better than others in infinitely iterated games. One of these approaches relies on the concept of  mixed strategy. In a game with several possible choices or moves, a player has a mixed strategy if he has a probability profile associated to all the possible moves of the game. Based on the concept of mixed strategy, the replicator dynamics approach, [9] and [10],  postulates  an evolution equation  for the probability profiles of each player's move. This evolution equation implies a precise type of rationality of the players, and the mixed strategy concept has a subjacent
infinite memory associated to the choices of the players. 

The formal construction of game theory depends on the relation between players and from whom they
receive their payoffs. For example, we can formalize a two-player game in such a way that the  payoffs won
by one player are the losses of the other, [11]. Another approach is to consider that the  player's payoffs are obtained from external sources. Our construction applies to the second case and  applies to games describing the global behaviour of systems from economy, sociology and evolutionary biology.

The aim of  this paper is to derive the mean payoffs of the 'representative players'  of a game, and to formulate the problem of deciding which policy performs better than another in  iterated non-cooperative games.  

This paper is organized as follows. In Section \ref{s2}, 
we introduce some of the definitions that will be used along this paper, 
and we analyse and interpret iterated non-cooperative games from the
point of view of dynamical system theory. In order to evaluate games and deterministic strategies with finite memory, we take the point of view of uniform ensembles of statistical mechanics and we
introduce the concept of representative ensemble of a game. In this context,  the players of the infinite set of games are substituted by the 'representative agents' of the game. 
 
In Section  \ref{s3}, we consider the case of a uniform ensemble of games, where in each subensemble we have two players playing the same game. The mean value of the payoffs per move taken over the uniform ensemble is calculated, and gives information about the performance of a game. In Section  \ref{s4}, we evaluate the performance of deterministic strategies with finite memory length.  In the case where in each subensemble a player has a deterministic strategy and the other 
makes his choices with equal probabilities, we calculate the ensemble averages of the payoffs per move.
The main results of sections  \ref{s3} and  \ref{s4} are summarized in Theorems  \ref{t1} and \ref{t2}, and  Corollary \ref{c1}. In particular, we show that if  one of the players has a generalized tit-for-tat policy,
the mean payoff per move of both players is the same. Therefore, generalized tit-for-tat is the best policy against exploitation.

In Section  \ref{s5}, we consider the case where the opponent players have deterministic policies, within the same memory class. In this case, the game dynamics is a deterministic process, and the mean payoffs per move depend on the 
initial moves of both players and on the policy functions of both players. 
Comparing all the possible deterministic strategies
with memory length $1$, we prove that, in dilemmatic games, the generalized tit-for-tat policy together with the condition of not being the first to defect, leads to the highest possible mean payoffs per move for the players.

In Section  \ref{s6}, we apply the formalism developed in this paper to the Prisoner's Dilemma and to the Hawk-Dove games, and we analyse their state space structure. Finally, in Section  \ref{s7}, we summarize the main conclusions of the paper.

\section{Formalism and definitions}\label{s2}

We take a two-player game  with two possible choices or moves --- two pure strategies. At times $n\ge 1$, each player chooses, independently of the other, one of the two possible pure strategies. These pure strategies are represented by the symbols '0' and '1'. We denote by ${\bf S}=\{0,1\}$ the set of  pure strategies, and by $P$ and $Q$ the two players. After a move, each player owns a profit
or payoff that is dependent of the opponent move. The payoff matrices of the game are:
$$
A=\left(\begin{array}{cc}
A_{00}&A_{01} \\ A_{10}&A_{11}
 \end{array} \right)\, ,\quad 
B=\left(\begin{array}{cc}
B_{00}&B_{01} \\ B_{10}&B_{11}
 \end{array} \right)
$$
where the payoff of $P$ is $A_{ij}$ if  player $P$ plays $i$ and $Q$ plays $j$. In the same move, $Q$ has payoff $B_{ij}$.  If each player makes its choice
independently of the other, we are in the context of  non-cooperative games. If $B=A^T$,  the two-player game is  symmetric. In the following, we analyze 
only the case of symmetric and non-cooperative games.

In a two-player symmetric game, we say that a pure strategy $i\in {\bf S}$ is do\-mi\-nant, [12], if
$A_{ij}\ge A_{kj}$, for every $j=0,1$, and  $A_{ij}> A_{kj}$ for some $j$ and $k\not= i$.
For example, the symmetric and non-cooperative games with payoff matrices,
$$
A_1=
\left(\begin{array}{cc} 3&0 \\ 5&1 \end{array} \right)
\, , \quad 
A_2=\left(\begin{array}{cc} 1&0 \\ 5&3 \end{array} \right)
$$
have '1'  as dominant strategy ($A_{10}>A_{00}$  and $A_{11}>A_{01}$).  In the first game, if   the two
players choose both the dominant strategy '1', their payoffs is $1$. In the
second game, the payoff of each players is $3$, and the dominant strategy is the right choice for both players. However, as $A_{00}>A_{11}$ for the first game, if  both players choose the non-dominant strategy, their individual payoffs per
move is higher when compared with the choice of the dominant strategy by both
players. 

These two examples suggest the following definition: A symmetric two-player game  is dilemmatic, if either,
$$
A_{10}>A_{00}>A_{11}>A_{01} \quad \hbox{and}\quad  2A_{00}>A_{01}+A_{10}\eqno(2.1)
$$
or,
$$
A_{01}>A_{11}>A_{00}>A_{10}\quad \hbox{and}\quad 2A_{11}>A_{10}+A_{01}\eqno(2.2)
$$
where the second inequalities in (2.1) and (2.2) have been introduced in order to favour the non-dominant strategy.

In the first case of a dilemmatic game, (2.1), the strategy '1' is dominant. In the second
case,  (2.2), '0' is the dominant strategy. If both players 
choose the dominant strategy in one move, they get  smaller payoffs than the ones they could have obtained if both had chosen the non-dominant strategy. 

In an iterated game with a fixed payoff matrix, players are always playing the same game, and their payoffs
accumulate. Therefore, a two-player iterated game is described by the two sequences of pure strategies of each player,
$$
\begin{array}{l}
{\bf \mu}=(\mu_1,\mu_2,\ldots ,\mu_n,\ldots )\\
{\bf \sigma}=(\sigma_1,\sigma_2,\ldots ,\sigma_n,\ldots ) 
\end{array}
\eqno(2.3)
$$
where $\mu_n$ and $\sigma_n$ represent the choices of the players $P$ and $Q$, respectively, at discrete time $n\ge 1$, and $\mu_n,\sigma_n\in {\bf S}$.  The sequences (2.3), completely specify the accumulated payoffs of both players. We  call ${\bf \mu}$ and ${\bf \sigma}$  the game record sequences.
In an infinitely iterated game, the accumulated payoff of the players can be infinite. The mean payoffs per move are always finite and, for a symmetric game, they are given by,  
$$
\begin{array}{l}
G_p=\lim_{n\to\infty}{1\over n}\sum_{i=1}^n A_{\mu_i\sigma_i}\\
G_q=\lim_{n\to\infty}{1\over n}\sum_{i=1}^n A_{\sigma_i\mu_i} 
\end{array} \eqno(2.4)
$$
where $G_p$ and $G_q$ are the mean payoffs of players $P$ and $Q$, respectively.

An example of a symmetric, non-cooperative, and dilemmatic game is the Prisoner's Dilemma game. In this game, we have two players
with two possible  pure strategies, '0' and '1', and we have chosen the payoff matrix,
$$
A=\left(\begin{array}{cc} 3&-9 \\ 11&-5 \end{array} \right)
\eqno(2.5)
$$
As,   $A_{10}>A_{00}>A_{11}>A_{01}$ and $2A_{00}>A_{01}+A_{10}$,
the Prisoner's Dilemma game is dilemmatic with '1' as dominant pure strategy. The pure strategy '0' corresponds to cooperation and the pure strategy '1' to defection. For a discussion about the importance of dilemmatic games and the Prisoner's Dilemma game, see the discussion in Axelrod, [3].

Following Neumann and Morgenstern [1],  a  strategy  or policy is a set of rules that tells each participant how to behave in every situation which may arise.
The only sources of information available to the players is the set
of all possible moves, their possible payoffs, and the history of the previous moves of both players. To describe 
a rule of decision, policy, or strategy we can adopt the Neumann-Morgenstern view where  
a rule of decision is specified through the knowledge of a function of the $m$ previous moves.

{\bf Deterministic  strategy}: In an iterated two-player game, with game records
${\bf \mu}$ and ${\bf \sigma }$ for players $P$ and $Q$, respectively, a rule of decision or a deterministic strategy with memory length $m\ge 1$ for player $P$  is a function $f:{\bf S}^{m}\to {\bf S}$ such that,
$$
\mu_i =f(\sigma_{i-m},...,\sigma_{i-1})\eqno(2.6)
$$
for every $i>m$.
Analogously, player $Q$ has a deterministic strategy with memory length $n\ge 1$, if there exists a function $g:{\bf S}^{n}\to {\bf S}$ such that,
$$
\sigma_i =g(\mu_{i-n},...,\mu_{i-1})
$$
for every $i>n$.
\bigskip

In the following,  deterministic policy and deterministic strategy
have the same meaning. In some game theory texts, the word 'strategy' refers to
'pure strategy', an element of the set ${\bf S}=\{0,1\}$,  and in other contexts if refers 
to policies, as in 'tit-for-tat strategy'.

By definition, the outcome of a player's choice or move at time $i\ge m+1$ is  determined by a finite number of previous moves of the other.
In  general, we can  take the
functions $f:{\bf S}^{m+n}\to {\bf S}$ and $g:{\bf S}^{r+s}\to {\bf S}$, and set,
$$
\begin{array}{l}
\mu_i =f(\sigma_{i-m},...,\sigma_{i-1},\mu_{i-n},...,\mu_{i-1})\\
\sigma_i =g(\mu_{i-r},...,\mu_{i-1},\sigma_{i-s},...,\sigma_{i-1}) 
\end{array}
$$
In the following we will only analyze  the case  where each player's
choice depends on a finite number of  previous moves of the other, and $m=n$.

For example,  adopting the definition of the tit-for-tat (TFT) strategy given in the introduction, a TFT strategy
with memory length $m=1$ is described by the boolean identity function $f:{\bf S}\to {\bf S}$, defined by,
$$
f(0)=0\qquad\hbox{and}\qquad f(1)=1
$$

{\bf Generalized tit-for-tat strategy (GTFT)}: We say that $f:{\bf S}^{m}\to {\bf S}$ is a generalized tit-for-tat strategy with memory length $m$, if the number of  '0' and '1' in $f({\bf \sigma}^{(m)})$, when  ${\bf \sigma}^{(m)}$ runs over  the set ${\bf S}^{m}$, are equal. More formally, $f:{\bf S}^{m}\to {\bf S}$ is a generalized tit-for-tat  strategy with memory length $m$, if,
$$
\#\{{\bf \sigma}^{(m)}\in {\bf S}^{m}:f({\bf \sigma}^{(m)})=0\}=
\#\{{\bf \sigma}^{(m)}\in {\bf S}^{m}:f(\sigma^{(m)})=1\}
$$
where ${\bf \sigma}^{(m)}=(\sigma_1,\ldots , \sigma_m)$, and $\sigma_i\in {\bf S}$.
\bigskip

To solve a game it is meant to find a procedure to determine for each player's choice 
which is the most favourable result 
([1], [11] and [13]). In this context, the concepts of mixed strategy and equilibrium state of a game are
fundamental tools in game theory.

A mixed strategy is a collection of probabilities associated to each player and 
its pure strategies. The players $P$ and $Q$ have mixed strategies $s_p=(s_{0p},s_{1p})$ and $s_q=(s_{0q},s_{1q})$, if each player plays pure strategy $i$ with 
probability $s_{i*}$. Obviously, $s_{0*}+ s_{1*}=1$. 

In a symmetric game with two pure strategies and mixed strategies $s_p$ and $s_q$ for players $P$ and $Q$,
respectively, the mean payoffs per move of players $P$ and $Q$ are,
$$
\begin{array}{ll}
P:\quad E(s_p|s_q)&=s_{0p}(s_{0q}A_{00}+s_{1q}A_{01})+s_{1p}(s_{0q}A_{10}+s_{1q}A_{11})\\
&=s_{0p}s_{0q}A_{00}+s_{0p}s_{1q}A_{01}+s_{1p}s_{0q}A_{10}+s_{1p}s_{1q}A_{11}\\
Q:\quad E(s_q|s_p)&=s_{0q}(s_{0p}A_{00}+s_{1p}A_{01})+s_{1q}(s_{0p}A_{10}+s_{1p}A_{11})\\
&=s_{0q}s_{0p}A_{00}+s_{0q}s_{1p}A_{01}+s_{1q}s_{0p}A_{10}+s_{1q}s_{1p}A_{11}
\end{array}
\eqno(2.7)
$$

The time evolution  of a game with  mixed  strategies $s_p$ and $s_q$  
can be seen as a stochastic processes with two independent random variables $X$ and
$Y$. The random variables $X$ and $Y$, associated to players $P$ and
$Q$, respectively, have mean values given by (2.7). More precisely, $X$ can assume
the values  $A_{00}$, $A_{01}$, $A_{10}$, $A_{11}$ with probabilities $s_{0p}s_{0q}$, $s_{0p}s_{1q}$, $s_{1p}s_{0q}$ and $s_{1p}s_{1q}$, respectively.
Analogously, $Y$ takes values in the same set, with probabilities: 
$s_{0q}s_{0p}$, $s_{0q}s_{1p}$, $s_{1q}s_{0p}$ and $s_{1q}s_{1p}$.
Therefore,
the deviations from the mean payoffs per move of the players, or the fluctuations from the mean values, are characterised by the variances,
$$
\begin{array}{ll}
\sigma^2_p(s_p|s_q)&=s_{0p}s_{0q}(A_{00}-E(s_p|s_q))^2+
s_{0p}s_{1q}(A_{01}-E(s_p|s_q))^2\\
&+s_{1p}s_{0q}(A_{10}-E(s_p|s_q))^2+
s_{1p}s_{1q}(A_{11}-E(s_p|s_q))^2\\
\sigma^2_q(s_q|s_p)&=s_{0q}s_{0p}(A_{00}-E(s_q|s_p))^2+
s_{0q}s_{1p}(A_{01}-E(s_q|s_p))^2\\
&+s_{1q}s_{0p}(A_{10}-E(s_q|s_p))^2+
s_{1q}s_{1p}(A_{11}-E(s_q|s_p))^2
\end{array}
\eqno(2.8)
$$
In general, $E(s_p|s_q)\not=E(s_q|s_p)$, but, by a straightforward calculation,  $\sigma^2_p(s_p|s_q)= \sigma^2_q(s_q|s_p)$. 

Imposing the condition $E(s_p|s_q)=E(s_q|s_p)$ in (2.7), a game or a mixed
strategy is equalitarian, if either, $A_{01}=A_{10}$, or $s_p=s_q$.

\begin{figure}[ht]
\centerline{\includegraphics[width=5.0 true cm]{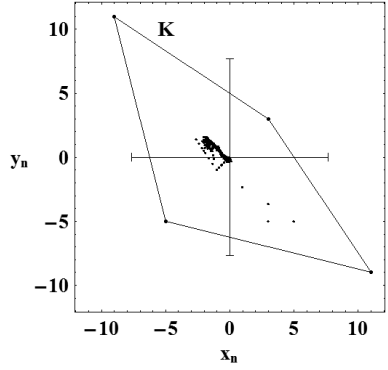}}
\caption{State space ${\mathcal K}$ of the Prisoner's Dilemma game with payoff matrix (2.5), where $x_n$ and $y_n$ are the mean payoffs per move of players $P$ and $Q$, as calculated from (2.9).
The dots represent the mean payoff per move of an instance of the iterated game with both players choosing strategies '0' and '1'  with equal probabilities ($s_p=s_q=1/2$).  By (2.7), the equilibrium of the game is the point $(0,0)$, and the standard deviations of the mean payoffs of the players are $\sigma_p=\sigma_q=\sqrt{59}=7.68$, calculated from (2.8). \label{fig1}}
\end{figure}

To characterise the dynamics of an iterated symmetric game, we  introduce the concept of phase
or state space of a game.
The state space of a two-player game is the convex closure of the points $(A_{00},A_{00})$, $(A_{11},A_{11})$, $(A_{01},A_{10})$ and $(A_{10},A_{01})$,  in the two-dimensional space of the payoffs of players $P$ and $Q$. Let us denote by ${\mathcal K}$ the state space of  a game. As $s_{0p}, s_{0q}\in [0,1]$,  
then $(E(s_p|s_q),E(s_q|s_p))\in {\mathcal K}$. In Fig. \ref{fig1}, we show the state space ${\mathcal K}$ for the Prisoner's Dilemma game with payoff matrix (2.5). 

In an iterated game,  the mean initial (at time $n=1$) payoffs per move of both players  is  $(x_1,y_1)=(A_{\mu_1\sigma_1},A_{\sigma_1\mu_1})\in {\mathcal K}$. By (2.4) and after $n+1$
moves, the mean payoffs per move of both players is,
$$\begin{array}{ll}
(x_{n+1},y_{n+1})&=({1\over {n+1}}\sum_{i=1}^{n+1} A_{\mu_i\sigma_i},{1\over {n+1}}\sum_{i=1}^{n+1} A_{\sigma_i\mu_i})
\\
&=({n\over {n+1}}x_{n}+{1\over {n+1}}A_{\mu_{n+1}\sigma_{n+1}},{n\over {n+1}}y_{n}+{1\over {n+1}}A_{\sigma_{n+1}\mu_{n+1}})\in {\mathcal K}
\end{array}
\eqno(2.9)
$$
and the iterated two-player game is dynamically described by a (non-deterministic) one-to-many map, $\beta:{\mathcal K}\to {\mathcal K}$, [14]. The equilibrium point or equilibrium solution of a game is the point $\lim_{n\to \infty} (x_n,y_n)$.

For a given mixed strategy profile $s_p$ and $s_q$ of players $P$ and $Q$, the ite\-ra\-ted two-player game has the equilibrium point,
$(E(s_p|s_q),E(s_q|s_p))\in {\mathcal K}$. As $s_{0p},s_{0q}\in[0,1]$, the set of equilibrium states of the map $\beta:{\mathcal K}\to {\mathcal K}$ span all the state space of  a game.

For example, in a two-player game with the mixed strategy profiles $s_p=(1/2,1/2)$ and 
$s_q=(1/2,1/2)$, by (2.7), the equilibrium point of the game is,
$$
({1\over 4}(A_{00}+A_{01}+A_{10}+A_{11}),{1\over 4}(A_{00}+A_{01}+A_{10}+A_{11}))
\eqno(2.10)
$$
By (2.8),  the fluctuations around the equilibrium are,
$$
\begin{array}{ll}
&\sigma^2_p((1/2,1/2)|(1/2,1/2))=\sigma^2_q((1/2,1/2)|(1/2,1/2))\\
&={1\over 4^2}\left(3A_{00}^2+3A_{01}^2+3A_{10}^2+3A_{11}^2\right.\\
&\left. -2A_{00}(A_{01}+A_{10}+A_{11})-2A_{11}(A_{01}+A_{10}) -2 A_{01}A_{10}\right) 
\end{array}
\eqno(2.11)
$$
In Fig. \ref{fig1}, we represent several iterates of the map $\beta$ for the Prisoner's Dilemma game with payoff matrix (2.5), and mixed strategies $s_p=s_p=(1/2,1/2)$. In the limit $n\to \infty$, $(x_n,y_n)\to (0,0)$. The fluctuations from equilibrium
have standard deviations $\sigma_p=\sigma_q=\sqrt{59}=7.68$.

A mixed strategy  is a strict Nash equilibrium solution of a game if $P$ and
$Q$ maximizes their payoffs per move independently of each other.  
A mixed strategy  is a  Nash bargain equilibrium solution of a game if $E(s_p|s_p)$ is a maximum.

In the case of the of  the Prisoner's Dilemma game with payoff matrix (2.5), by (2.7), we have,
$$
\begin{array}{ll}
P&:\quad E(s_p|s_q)=-5-4s_{0p}+16s_{0q}-4s_{0p}s_{0q}\\
Q&:\quad E(s_q|s_p)=-5-4s_{0q}+16s_{0p}-4s_{0p}s_{0q} 
\end{array}
\eqno(2.12)
$$
Maximizing $E(s_p|s_q)$ in order to $s_{0p}$ and $E(s_q|s_p)$ in order to $s_{0q}$, 
the strict Nash equilibrium of the game is obtained when both player choose the mixed strategy $(s_{0p},s_{1p})=(s_{0p},s_{1p})=(0,1)$.  In this case, the Nash
equilibrium state  of the map $\beta:{\mathcal K}\to {\mathcal K}$ is the point $(-5,-5)\in {\mathcal K}$. The 
Nash bargain solution of the game is obtained from (2.12) with $s_p=s_q$, and is the point $(3,3)\in {\mathcal K}$, Fig. \ref{fig1}. As both Nash solutions correspond to the choices of pure strategies with probability 1, by (2.8), the fluctuations of the iterated game have zero standard deviations.
In general,  a n-person non-cooperative game has
always a strict Nash equilibrium, [13].  

The choice of a mixed strategy profile for a game has the advantage that the iterates of the map 
$\beta:{\mathcal K}\to {\mathcal K}$ converges to the equilibrium solution $E(s_p|s_q)$. However, the choice of a mixed strategy profile implies that both players have infinite memory, which, in real situations, 
is difficult or even impossible to  fulfil.

On the other hand, in some game theory approaches describing the global behaviour of
economic, social and evolutionary systems, there are a large number of agents or players in mutual interaction. These individual agents interact with the same rules and
can also change partners along time. 
These situations are difficult to interpret under the infinite memory hypothesis, implicitly associated to the concept of mixed strategies.

Following this point of view, 
to evaluate a non-cooperative and symmetric game  and their possible deterministic strategies (short memory), we adopt the  point of view of the statistical ensembles of statistical mechanics. 
We suppose first that we have an infinite system composed by independent subensembles, where in
each subensemble we have two players playing the same game with payoff matrix $A$. 
We call this ensemble of independent games the uniform ensemble ([15, pp. 56]) of the game.
This uniform ensemble is characterized by the payoff matrix $A$, and the
players $P$ and $Q$ are the representative agents of the ensemble of the game. 

In each subensemble, a game with payoff matrix $A$ is played,  and subensembles are characterized by the mean  payoffs per move of both players. The global properties of the game will be described by the mean payoffs per move averaged over all the subensembles.
We say that the representative players $P$ and $Q$ of the game have mean payoffs per move 
${\bar G}_p$ and ${\bar G}_q$, respectively, where the average is taken over all the subensembles.  

To evaluate a game, we first consider  that each player chooses its pure strategies with equal probabilities, and each subensemble is characterized by the two sequences of pure strategies $\mu$ and $\sigma$. The properties
of the game are determined by ${\bar G}_p$ and ${\bar G}_q$. 



To evaluate a deterministic strategy or policy,  we consider that in each subensemble game,
$P$ plays with the deterministic policy $f$, and $Q$ has a game record  
${\bf \sigma}\in  {\bf S}^{\hbox{\N}}$. Defining an ensemble probability density function $\rho_q$ for the occurrence of game record ${\bf \sigma}$ for player $Q$, the ensemble of games
will by characterized by the mean payoff per move of both players averaged over the set of all allowed sequences ${\bf \sigma}\in  {\bf S}^{\hbox{\N}}$ with probability measure $\rho_q$. These averages depend on $f$ and $\rho_q$, and we can compare the performance of a policy with the case where the players have no policies.
Within the same memory class, we use these ensemble averages to compare the mean payoffs for  different policies.

When both players have a deterministic policy, the mean payoffs per move of the players depend on the finite number of initial conditions of the game.



\section{The uniform ensemble of a game}\label{s3}

We consider an ensemble of subsystems, where in each subsystem there are two players
playing the same game. We denote the game record of players $P$ and $Q$ by
${\bf \mu}$ and ${\bf \sigma}$, respectively. As ${\bf \mu}$ and ${\bf \sigma}$ are infinite sequences of '0' and '1', we can identify ${\bf \mu}$ and ${\bf \sigma}$ as real numbers in the interval $[0,1]$ through,
$$
x=\sum_{i=1}^\infty {\mu_i\over 2^i}\  ,\qquad
y=\sum_{i=1}^\infty {\sigma_i\over 2^i}\eqno(3.1)
$$
Relations (3.1) define  a map  
$\phi:{\bf S}^{\hbox{\N}}\rightarrow [0,1]$. The map $\phi$ is an isomorphism, except when $(\sigma_1,\sigma_2,...)$ or $(\mu_1,\mu_2,...)$ represents dyadic rational numbers,
[16]. As the set of dyadic rationals has zero Lebesgue measure, the infinite sequence of moves of both players  can be represented, almost everywhere, by two  real numbers  $x,y\in[0,1]$. Therefore, the interval $[0,1]$ is naturally
the space of game records.
 
Making this identification between  game records and real numbers,
in each subensemble game, by (2.4), the mean payoffs per move of the players are,
$$
\begin{array}{ll}
G_p({\bf \mu},{\bf \sigma})&=\lim_{n\to\infty}{1\over n}\sum_{i=1}^n A_{\mu_i\sigma_i}:=G_p(x,y) \cr
G_q({\bf \mu},{\bf \sigma})&=\lim_{n\to\infty}{1\over n}\sum_{i=1}^n A_{\sigma_i\mu_i}:=G_q(x,y)
\end{array} 
\eqno(3.2)
$$
where $x,y\in[0,1]$.

As each subensemble game is independent of the other, and each player's move is independent of the history of the game, we can assign ensemble  probability density functions to the game records. Let $\rho_p(x)$ and $\rho_q(y)$ be the ensemble probability density functions of game records of the representative players $P$ and $Q$, respectively. For example, $\rho_p(x) dx$ is the probability 
of finding a subensemble with player $P$ with a game record in an interval of length $dx$ centred around $x$.

Assuming further that all the game records are equally probable, $\rho_p(x)=1$ and 
$\rho_q(y)=1$, the ensemble averages of the mean payoffs per move are,
$$
\begin{array}{ll}
{\bar G}_p&=\int_0^1\int_0^1 G_p(x,y)\rho_p(x)\rho_q(y) dx\, dy =\int_0^1\int_0^1 G_p(x,y) dx\, dy\\
{\bar G}_q&=\int_0^1\int_0^1 G_q(x,y)\rho_p(x)\rho_q(y) dx\, dy =\int_0^1\int_0^1 G_q(x,y) dx\, dy
\end{array}
\eqno(3.3)
$$

To characterize the statistical ensemble of a non-cooperative and symmetric game with payoff matrix $A$, we now calculate the integrals in (3.3).
We consider the sequences of functions,
$$
\begin{array}{ll}
G_p^n({\bf \mu},{\bf \sigma})&={1\over n}\sum_{i=1}^n A_{\mu_i\sigma_i} \\
G_q^n({\bf \mu},{\bf \sigma})&={1\over n}\sum_{i=1}^n A_{\sigma_i\mu_i}\end{array} 
\eqno(3.4)
$$
As $n\to\infty$, $G_p^n({\bf \mu},{\bf \sigma})\to G_p({\bf \mu},{\bf \sigma})$, and $G_q^n({\bf \mu},{\bf \sigma})\to G_q({\bf \mu},{\bf \sigma})$. In the sense of Lebesgue integration,
the integrals in (3.3) can be calculated as the limits of the integrals of
the functions
$G_p^n({\bf \mu},{\bf \sigma})$ and $G_q^n({\bf \mu},{\bf \sigma})$.

Let us first take $n=1$. By (3.2), (3.3) and (3.4), we have,
$$
\begin{array}{ll}
{\bar G}_p^1& =\int_0^1\int_0^1 A_{\mu_1\sigma_1} dx\, dy\\
{\bar G}_q^1&=\int_0^1\int_0^1 A_{\sigma_1\mu_1} dx\, dy
\end{array}
\eqno(3.5)
$$
where $\mu_1$ and $\sigma_1$ are the first digits in the binary developments of $x$ and $y$, both in the interval $[0,1]$. Therefore, the functions 
$A_{\mu_1\sigma_1}\equiv A_{\mu_1\sigma_1}(x,y)$ and $A_{\sigma_1\mu_1}\equiv A_{\sigma_1\mu_1}(x,y)$ are piecewise constant in the unit square, and the integrals in (3.5) are straightforwardly evaluated to,
$$
\begin{array}{ll}
{\bar G}_p^1& =\int_0^1\int_0^1 A_{\mu_1\sigma_1} dx\, dy
={1\over 2^2}(A_{00}+A_{01}+A_{10}+A_{11})\\
{\bar G}_q^1&=\int_0^1\int_0^1 A_{\sigma_1\mu_1} dx\, dy
={1\over 2^2}(A_{00}+A_{01}+A_{10}+A_{11})
\end{array} 
\eqno(3.6)
$$
Note that, the functions $A_{\mu_1\sigma_1}(x,y)$ and $A_{\sigma_1\mu_1}(x,y)$ are
piecewise constant functions from $[0,1]\times [0,1]$
to the set $\{A_{00},A_{01},A_{10},A_{11}\}$.

In general, by (3.4) and (3.3),
$$
\begin{array}{ll}
{\bar G}_p^{n+1}& ={n\over n+1}{\bar G}_p^n+{1\over n+1} \int_0^1\int_0^1 A_{\mu_{n+1}\sigma_{n+1}}dx\, dy\\
{\bar G}_q^{n+1}&={n\over n+1}{\bar G}_q^n+{1\over n+1} \int_0^1\int_0^1 A_{\sigma_{n+1}\mu_{n+1}}dx\, dy
\end{array} 
\eqno(3.7)
$$
The functions $A_{\mu_{n}\sigma_{n}}\equiv A_{\mu_{n}\sigma_{n}}(x,y)$ and $A_{\sigma_{n}\mu_{n}}\equiv A_{\sigma_{n}\mu_{n}}(x,y)$ are piecewise constant and assume the constant
values $A_{00}$, $A_{01}$, $A_{10}$ and $A_{11}$ in squares of side $1/2^{n}$.
As, for each pair of indices $(\sigma_{n},\mu_{n})$, the domain where $A_{\sigma_{n}\mu_{n}}(x,y)$ is constant is composed
by $2^{2(n-1)}$ disjoint squares,  we have, 
$$
\begin{array}{ll}
\int_0^1\int_0^1 A_{\mu_{n}\sigma_{n}}dx\, dy&=\int_0^1\int_0^1 A_{\sigma_{n}\mu_{n}}dx\, dy\\
&={2^{2(n-1)}\over 2^{2n}}(A_{00}+A_{01}+A_{10}+A_{11})\\
&={1\over 2^2}(A_{00}+A_{01}+A_{10}+A_{11})
\end{array}
\eqno(3.8)
$$
Introducing (3.8) into (3.7), by induction, and taking the limit 
$n\to \infty$, we obtain the values of the ensemble average of the mean
payoffs per move of each player:

\begin{theorem}\label{t1}  We consider an ensemble of non-cooperative and
symmetric two-player game,
where in each subensemble we have  two players  making their choices with equal probabilities. Assume that each player's move is independent of the history of the game and that the ensemble probability density functions of each representative player are uniform in the interval $[0,1]$ of the game records. Then, the mean payoffs per move of the representative players of the game are equal and are
given by,
$$
{\bar G}_p={\bar G}_q={1\over 2^2}(A_{00}+A_{01}+A_{10}+A_{11})
$$
where the $A_{ij}$ are the entries of the payoff matrix.
\end{theorem}

In the uniform  statistical ensemble of a non-cooperative and symmetric game with all the players choosing their pure strategies with equal probabilities, the average payoff per move is equal to the average value of the entries of the payoff matrix $A$. 

These elementary results can be straightforwardly generalized to
 non-coope\-ra\-ti\-ve and
non-symmetric $n$-player games.

\section{Evaluating deterministic strategies}\label{s4}

To evaluate the performance of a deterministic strategy in an iterated game, we  first enumerate  the class of all the boolean functions $f:{\bf S}^m\rightarrow {\bf S}$, where ${\bf S}=\{0,1\}$. These boolean functions describe all the possible deterministic strategies. 

For each class of functions with memory length $m$, there are exactly $2^{2^m}$ different functions. To enumerate a deterministic policy function within a memory class $m$,
$f({{\bf \sigma}}^{(m)})$, where ${{\bf \sigma}}^{(m)}=(\sigma_1,\ldots ,\sigma_m)\in \{0,1\}^m$, we introduce an additional index  $n$. Within each memory class $m$,
each possible policy function will be denoted by $f_{m,n}$, where $n=\sum_{i=0}^{2^{m}-1}f_{m,n}({{\bf \sigma}}^{(m)}_i)2^i$ is the policy number,  ${{\bf \sigma}}^{(m)}_{i+1}={{\bf \sigma}}^{(m)}_i+(0,0,\ldots ,1)$,  ${{\bf \sigma}}^{(m)}_0=(0,0,\ldots ,0)$, and the "plus" symbols must be understood in the sense of binary arithmetic. For example,  in Table \ref{table1}, we show all  the possible deterministic policy functions with memory length $m=1$.

\begin{table}[h]
\centerline{
\begin{tabular}{@{}ccc@{}}
\hline
{} &{}&{}\\[-1.5ex]
{} & $f_{1,n}(0)$&$f_{1,n}(1)$\\[1ex]
\hline
{} &{} &{} \\[-1.5ex]
$f_{1,0}$& 0& 0\\[1ex]
$f_{1,1}$& 1& 0\\[1ex]
$f_{1,2}$& 0& 1\\[1ex]
$f_{1,3}$& 1& 1\\[1ex] 
\hline
\end{tabular}
}
\caption{Deterministic policy functions for $m=1$.}\label{table1}  
\end{table}

In this case, the deterministic TFT policy  corresponds to the boolean function $f_{1,2}$. The functions $f_{1,2}$ and $f_{1,1}$ are GTFT policies with memory length $m=1$.

Suppose now that the representative player $P$ of a game has a policy $f_{m,n}$ and the opponent player $Q$ can have  any
game sequence $\sigma=(\sigma_1,\sigma_2,...)$. Then, by (2.4), for the infinitely iterated game, the mean payoff  per move  for each player is,
$$
\begin{array}{ll}
G_p({\bf \sigma})&=\lim_{M\to\infty}{1\over M} \sum_{i=1}^M A_{\mu_i\sigma_i}\\
&=\lim_{M\to\infty}{1\over M-m} \sum_{i=m+1}^M A_{f_{m,n}(\sigma_{i-m},...,\sigma_{i-1}) \sigma_i}   \\
G_q({\bf \sigma})&=\lim_{M\to\infty}{1\over M-m} \sum_{i=m+1}^M A_{\sigma_i f_{m,n}(\sigma_{i-m},...,\sigma_{i-1})}
\end{array}
\eqno(4.1)
$$
and $G_p$ and $G_q$ are functions of ${\bf \sigma}=(\sigma_1,\sigma_2,...)$ and $f_{m,n}$.
In the first $m$ iterations of the game, the accumulated mean payoffs depend on the initial choices of the players. However, 
in the limit $M\to \infty $, the dependence on the  initial choices vanishes.

Let us take the infinite sequence $(\sigma_1,\sigma_2,...)\in {\bf S}^{\hbox{\N}}$ characterizing one of the possible outcomes of the choices of the player $Q$,
and define the real number,
$$
y=\sum_{i=1}^\infty {\sigma_i\over 2^i}\eqno(4.2)
$$
With this identification between infinite sequences of zeros and ones with real numbers in the interval $[0,1]$,
we write  the mean payoffs as,
$$
\begin{array}{l}
G_p({\bf \sigma})\equiv G_p(y;f_{m,n}):=P_{m,n}(y) \\
G_q({\bf \sigma})\equiv G_q(y;f_{m,n}):=Q_{m,n}(y)
\end{array}
\eqno(4.3)
$$

Let us suppose now that we are in framework of statistical mechanics and we have
an ensemble or collectivity of players $P$ and $Q$. In each subensemble of the collectivity, the player $P$ plays according to  strategy $f_{m,n}$ and $Q$ has some  game record $y\in[0,1]$. Suppose additionally that all the subensembles of the collectivity are independent.

As each member of the collectivity is independent of the others, we can assign  an ensemble density function $\rho_q(y)$ to the collectivity. The function  $\rho_q(y)$ is the probability density
of  the game record $y$ of  player $Q$.
If $\rho_q(y)=1$, all the game records of $Q$ are equally probable. 
The uniform ensemble of the game 
can then be characterized by the ensemble  mean payoffs per move and per player,
$$
\begin{array}{ll}
{\bar P}_{m,n}&=\int_0^1 P_{m,n}(y)\rho_q(y) dy =\int_0^1 P_{m,n}(y)  dy \\
{\bar Q}_{m,n}&=\int_0^1 Q_{m,n}(y)\rho_q(y) dy=\int_0^1 Q_{m,n}(y)  dy\end{array}
\eqno(4.4)
$$

\begin{figure}[ht]
\centerline{\includegraphics[width=12.0 true cm]{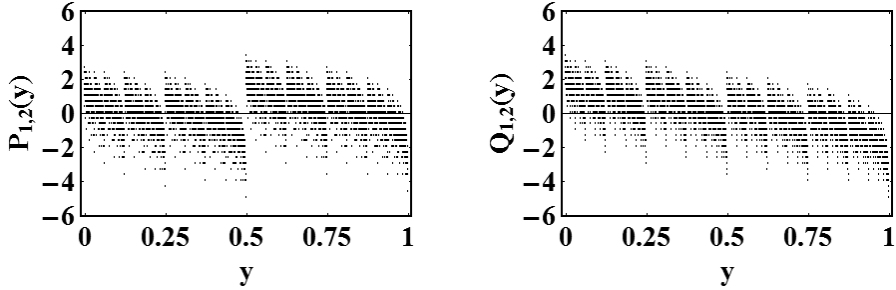}}
\caption{Mean payoffs per move $P_{1,2}(y)$ and
$Q_{1,2}(y)$ as a function of the game record $y\in[0,1]$ of the player $Q$, in the Prisoner's Dilemma game with payoff matrix (2.5). Player $P$ has TFT policy $f_{1,2}$.  \label{fig2}}
\end{figure}

In Fig. \ref{fig2}, we show the mean payoff functions $P_{m,n}(y)$ and
$Q_{m,n}(y)$ for the TFT policy $f_{1,2}$ and payoff matrix (2.5) of the Prisoner's Dilemma game.
These functions have been calculated numerically from (4.1), (4.2) and (4.3).

To calculate the mean payoffs ${\bar P}_{m,n}$ and ${\bar Q}_{m,n}$ given by (4.4),  we first approximate the functions $P_{m,n}(y)$
and $Q_{m,n}(y)$ by sequences of piecewise constant functions. 
By (4.1)-(4.3), we define the sequences of functions, $\{P_{m,n}^M(y)\}_{M\ge m+1}$ and $\{Q_{m,n}^M(y)\}_{M\ge m+1}$, as,
$$
\begin{array}{ll}
P_{m,n}^M(y)&={1\over M-m} \sum_{i=m+1}^M A_{f_{m,n}(\sigma_{i-m},...,\sigma_{i-1}) \sigma_i}  \\
Q_{m,n}^M(y)&={1\over M-m} \sum_{i=m+1}^M A_{\sigma_i f_{m,n}(\sigma_{i-m},...,\sigma_{i-1})} 
\end{array} 
\eqno(4.5)
$$
where  $(\sigma_1,\sigma_2,...)$ is the binary development of $y$.
For  $M\ge m+1$, $P_{m,n}^M(y)$ and $Q_{m,n}^M(y)$ are piecewise constant functions in the interval $[0,1]$, and, in the limit $M\to\infty$, they converge almost everywhere
to $P_{m,n}(y)$ an $Q_{m,n}(y)$, Fig. \ref{fig2}. In the sense of Lebesgue integrations, this implies that, 
$$
\begin{array}{l}
\lim_{M\to\infty}\int_0^1 P_{m,n}^M(y) dy ={\bar P}_{m,n}\\
\lim_{M\to\infty}\int_0^1 Q_{m,n}^M(y) dy={\bar Q}_{m,n}
\end{array} 
\eqno(4.6)
$$

Let us now calculate the integrals in (4.6).
For $M=m+1$, by (4.5), we have,
$$
\begin{array}{ll}
P_{m,n}^{m+1}(y)&= A_{f_{m,n}(\sigma_{1},...,\sigma_{m}) \sigma_{m+1}}\\
Q_{m,n}^{m+1}(y)&=A_{\sigma_{m+1}f_{m,n}(\sigma_{1},...,\sigma_{m}) }
\end{array} 
\eqno(4.7)
$$
As $(\sigma_1,\ldots ,\sigma_{m+1})$ represents the first $m+1$ terms of the binary development of $y$, the functions  in (4.7) assume constant values in subintervals of $[0,1]$ of length $1/2^{m+1}$. In each
of these subintervals, $P_{m,n}^{m+1}(y)$ and $Q_{m,n}^{m+1}(y)$ assume one of the four values: $A_{00}$, $A_{01}$, $A_{10}$ and $A_{11}$. 

Associated to each deterministic policy function $f_{m,n}$, we define the numbers,
$$
\begin{array}{ll}
n_{0}^{m,n}&=\#\{{\bf \sigma}^{(m)}\in\{0,1\}^m:f_{m,n}({\bf \sigma}^{(m)})=0 \}\\
n_{1}^{m,n}&=\#\{{\bf \sigma}^{(m)}\in\{0,1\}^m:f_{m,n}({\bf \sigma}^{(m)})=1\}
\end{array} 
\eqno(4.8)
$$
and, $n_{0}^{m,n}+n_{1}^{m,n}=2^m$.

Under these conditions, by  (4.7) and (4.8), we have,
$$
\begin{array}{ll}
\int_0^1 P_{m,n}^{m+1}(y)dy&={1\over 2^{m}} \left( n_{0}^{m,n} {(A_{00}+A_{01})\over 2}+  n_{1}^{m,n} {(A_{10}+A_{11})\over 2} \right) \\
&={1\over 2^{m+1}} \left( n_{0}^{m,n} (A_{00}+A_{01})+  n_{1}^{m,n} (A_{10}+A_{11}) \right) \\
\int_0^1 Q_{m,n}^{m+1}(y)dy&={1\over 2^{m+1}} \left( n_{0}^{m,n} (A_{00}+A_{10})+  n_{1}^{m,n} (A_{01}+A_{11}) \right)
\end{array}
\eqno(4.9)
$$

For $M>m+1$, by (4.5), we have,
$$
\begin{array}{ll}
P_{m,n}^{M+1}(y)&={M-m\over M-m+1} P_{m,n}^{M}(y)+ {1\over M+1-m}A_{f_{m,n}(\sigma_{M+1-m},...,\sigma_{M}) \sigma_{M+1}}   \\
Q_{m,n}^{M+1}(y)&={M-m\over M-m+1} Q_{m,n}^{M}(y)+ {1\over M+1-m}A_{\sigma_{M+1} f_{m,n}(\sigma_{M+1-m},...,\sigma_{M}) }
\end{array}
\eqno(4.10)
$$
and, as in (4.9),
$$
\begin{array}{ll}
\int_0^1 P_{m,n}^{M+1}(y) dy =&{M-m\over M-m+1}\int_0^1 P_{m,n}^{M}(y) dy \\
&+{1\over 2^{m+1} (M+1-m) } \left( n_{0}^{m,n} (A_{00}+A_{01})+  n_{1}^{m,n} (A_{10}+A_{11}) \right)\\
\int_0^1 Q_{m,n}^{M+1}(y) dy =&{M-m\over M-m+1}\int_0^1 Q_{m,n}^{M}(y) dy \\
&
+{1\over 2^{m+1}  (M+1-m) } \left( n_{0}^{m,n} (A_{00}+A_{10})+  n_{1}^{m,n} (A_{01}+A_{11}) \right)
\end{array} 
\eqno(4.11)
$$
By (4.9) and by induction from  (4.11),  we obtain,
$$
\begin{array}{ll}
\int_0^1 P_{m,n}^{M+1}(y) dy =&{1\over 2^{m+1}  } \left( n_{0}^{m,n} (A_{00}+A_{01})+  n_{1}^{m,n} (A_{10}+A_{11}) \right)\\
\int_0^1 Q_{m,n}^{M+1}(y) dy =&{1\over 2^{m+1}  } \left( n_{0}^{m,n} (A_{00}+A_{10})+  n_{1}^{m,n} (A_{01}+A_{11}) \right)
\end{array}
\eqno(4.12)
$$
As the integrals in (4.12) are independent of $M$, by (4.6), we have proved:

\begin{theorem}\label{t2}  We consider  an ensemble of non-cooperative and symmetric two-player games,
where in each subensemble we have  a  player $P$ playing with deterministic policy $f_{m,n}$, and a  player $Q$ making the choices
of pure strategies with equal probabilities. Then, the mean payoffs per move of the representative players
of the game  depend on the payoff matrix $A$
and on the strategy $f_{m,n}$, and the mean payoffs per move are,
$$
\begin{array}{ll}
{\bar P}_{m,n} =&{1\over 2^{m+1}  } \left( n_{0}^{m,n} (A_{00}+A_{01})+  n_{1}^{m,n} (A_{10}+A_{11}) \right)\cr
{\bar Q}_{m,n} =&{1\over 2^{m+1}  } \left( n_{0}^{m,n} (A_{00}+A_{10})+  n_{1}^{m,n} (A_{01}+A_{11}) \right)
\end{array} 
$$
where the $A_{ij}$ are the entries of the payoff matrix, 
$n_{0}^{m,n}=\#\{{\bf \sigma}^{(m)}\in\{0,1\}^m:f_{m,n}({\bf \sigma}^{(m)})=0\}$, and $n_{1}^{m,n}=\#\{{\bf \sigma}^{(m)}\in\{0,1\}^m :f_{m,n}({\bf \sigma}^{(m)})=1\}$.
\end{theorem}

This theorem  has a direct consequence. With the definitions of   Section \ref{s2}, a policy or strategy $f_{m,n}$ is equalitarian 
if the mean payoffs of the representative players are equal. Imposing the equality between ${\bar P}_{m,n}$ and ${\bar Q}_{m,n}$ in
Theorem \ref{t2}, we obtain,
$$
n_{0}^{m,n} (A_{01}-A_{10})+  n_{1}^{m,n} (A_{10}-A_{01})=0\eqno(4.13)
$$
From (4.13) it follows that a policy is equalitarian if either $n_{0}^{m,n}=n_{1}^{m,n}$ 
or, $A_{01}=A_{10}$. In the first case, we have the class of all GTFT policies, independently of the values of the entries of the payoff matrix $A$. 

If $A_{01}=A_{10}$, it follows from Theorem \ref{t2} and (4.13), that,
$$
{\bar P}_{m,n}={\bar Q}_{m,n}={n_{0}^{m,n}\over 2^{m+1}}(A_{00}-A_{11})+{1\over 2} A_{11}+{1\over 2} A_{01}\eqno(4.14)
$$
where we have introduced the relation $n_{1}^{m,n}=2^m-n_{0}^{m,n}$. 
Therefore, we have:

\begin{corollary}\label{c1} 
We consider an ensemble of non-cooperative and symmetric two-player games with payoff matrix $A$,
where in each subensemble we have  a  player $P$ playing strategy $f_{m,n}$, and a  player $Q$ making the choices
of pure strategies with equal probabilities.  Then the policy $f_{m,n}$ is equalitarian if either, it is GTFT or, $A_{01}=A_{10}$. Moreover, the payoffs per move of GTFT policies are given by,
$$
{\bar P}_{m,n} ={\bar Q}_{m,n}={1\over 2^2  } \left( A_{00}+A_{01}+ A_{10}+A_{11} \right)
$$
\end{corollary}

For example, in games with memory length $m=1$, independently of the payoff matrix $A$, the equalitarian strategies are
$f_{1,1}$ and $f_{1,2}$, both GTFT. From the point of view of the ensemble mean payoffs per move, all the GTFT strategies are equivalent to ensemble games where all player play randomly with equal probability.

We 
determine now the best policy for a player $P$ with an opponent
$Q$  choosing pure strategies with equal probabilities. 
By Theorem \ref{t2}, and with
$n_{1}^{m,n}=2^m-n_{0}^{m,n}$, we obtain,
$$
{\bar P}_{m,n}={n_{0}^{m,n}\over 2^{m+1}}(A_{00}+A_{01}-A_{11}-A_{10})+{1\over 2} (A_{11}+ A_{10})\eqno(4.15)
$$
Therefore, in the sense of ensemble average and for a given memory length $m$, the best policies for the player $P$ are the ones that maximise (4.15), for all the choices of the integers $n_{0}^{m,n}=0,\ldots ,2^m$.

\section{Both players have deterministic strategies}\label{s5}

When the two representative players $P$ and $Q$ have deterministic strategies within the same memory class, their game records become dependent of the first $m$ moves 
of the players. As we have two players and $2^{m}$ different initial
conditions for each player, for each choice of a pair of deterministic strategies, there are at most $2^{m+1}$  different payoffs per move 
for both players.  As there are $2^{2^m}$ different boolean functions
of memory length $m$, the maximum number of equilibrium states is,
$2^{2^{m+1}}\times 2^{m+1}$, which, for $m=1$, is $64$.

Let us analyse now in detail the case of memory length $m=1$.
If $\mu_1$ and $\sigma_1$ represent the choices for the first move of players $P$ and $Q$, and $P$ and $Q$ have policies $f\equiv f_{1,r}$ and $g\equiv f_{1,s}$, respectively, their game records are,
$$
\begin{array}{ll}
P:\ &(\mu_1,f(\sigma_1),f\cdot g(\mu_1),f\cdot g\cdot f(\sigma_1),\ldots )\\
Q:\ &(\sigma_1,g(\mu_1),g\cdot f(\sigma_1),g\cdot f\cdot g(\mu_1),\ldots )
\end{array} 
$$
where $f\cdot g(\mu_1)=f( g(\mu_1))$.
After a few moves, the game records become periodic. Therefore, the mean payoff per
move of each player can be calculated by the periodic sequences which depend on the initial moves and on the policies. For example, with $f\equiv f_{1,2}$ and $g\equiv f_{1,2}$, and initial moves   $\mu_1= 0$ and $\sigma_1=1$, we obtain the 
game records,
$$
\begin{array}{ll}
P:\ &(0,1,0,1,\ldots )\\
Q:\ &(1,0,1,0,\ldots   ) 
\end{array} 
$$
and the mean payoff per move of both players is ${\bar P}={\bar Q}=(A_{01}+A_{10})/2$. But for the initial moves $\mu_1= 0$ and $\sigma_1=0$, we have, ${\bar P}={\bar Q}=A_{00}$.

\begin{table}[hp]
{\footnotesize
\centerline{
\begin{tabular}{@{}ccccc@{}}
\multicolumn{5}{c}{$P:\, f_{1,0}$}\\
\hline
{}& {} & {} & {} & {} \\[-1.5ex]
${\mu_1 \choose\sigma_1}$& $Q:\, f_{1,0}$ & $Q:\, f_{1,1}$ & $Q:\, f_{1,2}$ & $Q:\, f_{1,3}$ \\[1ex]
\hline
{} &{} &{} &{} &{} \\[-1.5ex]
&$A_{00}$& $P:A_{01};\, Q: A_{10}$&  $A_{00}$& $P:A_{01};\, Q: A_{10}$\\[1ex]
\hline
\end{tabular}}}

\vspace{5pt}
{\footnotesize
\centerline{
\begin{tabular}{@{}ccccc@{}}
\multicolumn{5}{c}{$P:\, f_{1,1}$}\\
\hline
{}& {} & {} & {} & {} \\[-1.5ex]
${\mu_1 \choose\sigma_1}$& $Q:\, f_{1,0}$ & $Q:\, f_{1,1}$ & $Q:\, f_{1,2}$ & $Q:\, f_{1,3}$ \\[1ex]
\hline
{} &{} &{} &{} &{} \\[-1.5ex]
${0 \choose 0}$&$P: A_{10};\, Q: A_{01}$& ${A_{11}+A_{00}\over 2}$&$  {B\over 4}$& $P:A_{01};\, Q: A_{10}$\\[1ex]
${0 \choose 1}$&$P: A_{10};\, Q: A_{01}$&$ P:A_{01};\, Q: A_{10}$&  ${B\over 4}$& $P:A_{01};\, Q: A_{10}$\\[1ex]
${1 \choose 0}$&$P: A_{10};\, Q: A_{01}$& $P:A_{10};\, Q: A_{01}$&  ${B\over 4}$& $P:A_{01};\, Q: A_{10}$\\[1ex]
${1 \choose 1}$&$P: A_{10};\, Q: A_{01}$&  ${A_{11}+A_{00}\over 2}$&  ${B\over 4}$& $P:A_{01};\, Q: A_{10}$\\[1ex]
\hline
\end{tabular}}}

\vspace{5pt}
{\footnotesize
\centerline{
 \begin{tabular}{@{}ccccc@{}}
\multicolumn{5}{c}{$P:\, f_{1,2}$}\\
\hline
{}& {} & {} & {} & {} \\[-1.5ex]
${\mu_1 \choose\sigma_1}$& $Q:\, f_{1,0}$ & $Q:\, f_{1,1}$ & $Q:\, f_{1,2}$ & $Q:\, f_{1,3}$ \\[1ex]
\hline
{} &{} &{} &{} &{} \\[-1.5ex]
${0 \choose 0}$&$A_{00}$& ${B\over 4}$&$A_{00}$& $A_{11}$\\[1ex]
${0 \choose 1}$&$A_{00}$&$ {B\over 4}$&  ${A_{10}+A_{01}\over 2}$& $A_{11}$\\[1ex]
${1 \choose 0}$&$A_{00}$& ${B\over 4}$&  ${A_{10}+A_{01}\over 2}$& $A_{11}$\\[1ex]
${1 \choose 1}$&$A_{00}$&  ${B\over 4}$&  $A_{11}$& $A_{11}$\\[1ex]
\hline
\end{tabular}}}

\vspace{5pt} 
{\footnotesize
\centerline{
\begin{tabular}{@{}ccccc@{}}
\multicolumn{5}{c}{$P:\, f_{1,3}$}\\
\hline
{}& {} & {} & {} & {} \\[-1.5ex]
${\mu_1 \choose\sigma_1}$& $Q:\, f_{1,0}$ & $Q:\, f_{1,1}$ & $Q:\, f_{1,2}$ & $Q:\, f_{1,3}$ \\[1ex]
\hline
{} &{} &{} &{} &{} \\[-1.5ex]
&$P:A_{10};\, Q: A_{01}$& $P:A_{10};\, Q: A_{01}$&  $A_{11}$& $A_{11}$\\[1ex]
\hline
\end{tabular}  }}
\caption{Mean payoffs per move of players $P$ and $Q$ as a function of the initial move and  policies with memory length $m=1$. In the first and forth tables, the 
mean payoffs per move of player $Q$ are independent of
the initial move.
To simplify the notation, we have done $B=(A_{00}+A_{01}+A_{10}+A_{11})$. When an entry shows only one payoff, this payoff is the same for both players. The TFT strategy corresponds to the deterministic strategy function $ f_{1,2}$, and $f_{1,1}$ and $f_{1,2}$ are GTFT policies.}
\label{table2}  
\end{table}

In Table \ref{table2}, we show the mean payoffs per move and per player, for all the deterministic policies with memory length $m=1$ and all the possible four different initial moves of the players. 
Counting the different values in the entries in table, we conclude that, for $m=1$, the number of equilibrium states is $7$.
For a given game, the best strategy and initial conditions is obtained by analyzing the entries of 
Table \ref{table2}. Clearly, the best strategy depends on
the entries of the payoff matrix of the game.

In general, let $(\mu_1,\ldots ,\mu_m)$ and $(\sigma_1,\ldots ,\sigma_m)$ be the first $m$ moves of players $P$ and $Q$, respectively.
Suppose further that player $P$ and $Q$ choose the deterministic strategies $f_{m,n}$ and $g_{m,n}$, respectively. Iterating the game,
after some transient iteration the  game record sequences become periodic, and the  mean payoffs per move and per player are easily calculated. If $(\mu_{i+1},\ldots ,\mu_{i+p})$ and $(\sigma_{i+1},\ldots ,\sigma_{i+p})$, for some $i\ge 1$,  are the periodic patterns of period $p$ of the game
record sequences, the mean payoffs per move of the players are,
$$
\begin{array}{ll}
P:\ &{1\over p}(A_{\mu_{i+1}\sigma_{i+1}}+\ldots +A_{\mu_{i+p}\sigma_{i+p}})\\
Q:\ &{1\over p}(A_{\sigma_{i+1}\mu_{i+1}}+\ldots +A_{\sigma_{i+p}\mu_{i+p}})
\end{array}
$$
and these mean payoffs are the equilibrium states of the game.
For example, for $m=2$, we have at most $2^{2^{2+1}}\times 2^{2+1}=2048$ equilibrium states.

\section{Examples and policy analysis}\label{s6}

The formalism  introduced in the previous sections leads
to the evaluation of  policies for an iterated game with a given payoff matrix $A$. In this context, we can forget the role of players $P$ and
$Q$ and speak about the performance of the game, the performance of a deterministic strategy and the relative performance of two deterministic strategies.

We analyze now two examples, the Prisoner's Dilemma game and the Hawk-Dove game.

\subsection{The Prisoner's Dilemma}

In the Prisoner's Dilemma game  with payoff matrix (2.5), if
all players make their choices with equal probabilities, by Theorem \ref{t1}, the mean payoff   per move and per player  is zero. 
By Corollary 4.2, a player with a  GTFT policy against a player choosing its pure strategies randomly has also zero payoffs.
This includes the simplest case of the tit-for-tat policy. 

\begin{figure}[ht]
\centerline{\includegraphics[width=12.0 true cm]{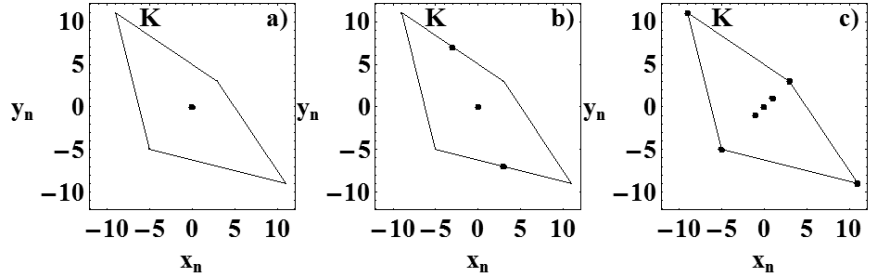}}
\caption{State space ${\mathcal K}$ and equilibrium solutions of the iterated Prisoner's Dilemma game. In a), all the players in the collectivity choose their strategies with equal probabilities. In b), we have three possible equilibrium states. In each subensemble game one player follows a deterministic strategy with memory length 
$m=1$, and the other chooses its pure strategies with equal probabilities. Depending on the adopted policy, we can have different equilibrium states.
In c), all the players have chosen a deterministic strategy with memory length $m=1$, and the game can have seven equilibrium states.  \label{fig3}}
\end{figure}

From the point of view of the non-deterministic map $\beta :{\mathcal K}\to {\mathcal K}$, the situation of Theorem \ref{t1} corresponds to the equilibrium solution $(0,0)\in {\mathcal K}$, Fig. \ref{fig3}a). 

In the case of Theorem \ref{t2} and for deterministic strategies with memory length $m=1$, there
are three equilibrium solutions for the  Prisoner's Dilemma game. 
These equilibrium solutions are: $(-3,7)\in {\mathcal K}$, $(0,0)\in {\mathcal K}$ and $(3,-7)\in {\mathcal K}$, Fig. \ref{fig1}b). So, in a uniform collectivity, the players  that choose the dominant strategy  have a better payoff, provided their partners choose their strategies with equal probabilities.
 
Suppose now that both representative players $P$ and $Q$ adopt a policy with memory length
$m=1$. Analysing the results of Table \ref{table2},  the best payoff per move for both players is obtained
when player $P$ and $Q$ play tit-for-tat and both choose the initial strategy   '0'. The best payoff per move is also obtained when one at least of the contenders  chooses '0' and the other plays according the tit-for-tat policy.  In the case of policies of memory length 
$m=1$,  the tit-for-tit policy forces cooperation. If one of the player plays tit-for-tat and the other player chooses another strategy, tit-for-tit ensures that the payoffs per move of both players are equal and the second player is not able to increase its payoff per move. If both players choose   a tit-for-tat policy, depending on the initial condition, we can have four different equilibrium states of the game, Table \ref{table2} and  Fig. \ref{fig1}c). In this case, two of them are the strict and the bargain Nash solutions. If one of the players chooses always
the strategy  '1', it corresponds to the deterministic policy $f_{1,3}$, and the outcome of the game against a
tit-for-tat corresponds to the Nash strict equilibrium of the game. The seven equilibrium solutions of the Prisoner's game are plotted in Fig. \ref{fig3}c).

If $P$ has to choose a policy against a player  $Q$ that plays its strategies with equal probabilities, by (4.15) and (2.5),  the best policy for $P$ is the one that maximizes,
$$
{\bar P}_{m,n}=\left( 3-12{n_{0}^{m,n}\over 2^{m+1}}\right)
$$
Therefore, in the Prisoner's Dilemma game, the best policy corresponds to
$n_{0}^{m,n}=0$, which corresponds to the policy function $f_{m,2^m-1}$. 
In this case, we have, ${\bar P}_{m,2^m-1}=3$ and ${\bar Q}_{m,2^m-1}=-7$, Fig. \ref{fig3}b).

A more detailed analysis of Fig. \ref{fig3}, shows that Nash bargain solutions 
and Nash strict solutions only exist when both players have deterministic policies. In the sense of ensemble averages,
Nash solutions are not equilibrium solutions of a game.

\subsection{The Hawk-Dove game}

The Hawk-Dove game has been introduced by Maynard-Smith and Price [17] as a game theoretical basic model  to describe animal conflicts. They have assumed two pure strategies: Hawk ('0') and Dove ('1').
A player chooses Hawk or '0' if he acts fiercely, and chooses 
Dove or '1' if he looks fierce and then retires. In the context of evolutionary biology, this game aims to explain the struggle for a territory whose payoff is related with the number of offsprings.
The payoff matrix of the Hawk-Dove game is,
$$
A=\left(\begin{array}{cc}{1\over 2}(r-c)&r \\ 0&{r\over 2}+\varepsilon
 \end{array} \right)
$$
where $r$ represents the reproductive value and $c$ is the cost of injury.
In this game the Hawk strategy is dominant, provided $c<r$ and $\varepsilon<r/2$. If $\varepsilon>0$ and $\varepsilon<r/2$, the Hawk-Dove game is also dilemmatic. Globally, the species has advantage if everybody acts Dove, which is the non-dominant strategy.

If all players choose their strategies with equal probability, by Theorem \ref{t1}, the mean payoff  per player and move is,
$$
{\bar P}={\bar Q}={1\over 2}(r-{1\over 4}c)+\varepsilon
$$ 
If  $c< 4r+8\varepsilon$,  ${\bar P}={\bar Q}>0$, the Hawk-Dove game shows advantage 
for the species. If $c\ge 4 r+8\varepsilon$, the cost of injury is too high and globally
the mean payoff per player and move is non-positive.

If  the representative players of the game choose a generalized tit-for-tat policy with memory length $m=1$, and $c<r$, both players have a positive mean payoff per move. If the players choose not being the first to play Hawk, they both obtain the highest mean payoffs per move. 

In the Hawk-Dove non-cooperative and symmetric game, the tit-for-tat policy or imitation of the adversary move implies a positive payoff for
both players, provided the cost of injury is not too high ($c< 4r+8\varepsilon$).

\section{Conclusions}\label{s7}

The dynamics of an iterated game is described by a
one-to-many map defined on a state space, [14]. Within this framework, 
the concept of mixed strategy  leads to the definition of the equilibrium  solution of a game. This equilibrium solution is obtained as
the limit of the iterates of a one-to-many map. In general, for a specific game,  the
equilibrium solutions associated to the set of all mixed strategies span the state space of the game. The concepts of strict Nash equilibrium and bargain solutions of a game are discussed within this framework.

In applications of game theory to economics, ethics, sociology, biology, physics, {\it etc.}, it is sometimes  easy to identify rules of behaviour and interactions between agents and to make guesses about payoffs. However, it is difficult to argue about the (infinite) memory of all the past choices of the  players, and to insure that opponent players remain the same during all the iterated game. Therefore, the way of evaluating a game, or a policy depends on the context in which the game
is considered.

In order to evaluate a game, we have introduced  the concept of representative ensemble of a game. This technique has been applied 
to the global evaluation of a game, without  any specific considerations about policies. In this evaluation, all the players make their choices of pure strategies with equal probabilities.
In this case, we have shown that the mean payoffs per move of the players are the mean value
of the entries of the payoff matrices of the game. 

To evaluate a deterministic policy with a finite memory length, we
have  calculated the  mean payoffs per move of the players, for the case where one of the players has a deterministic policy and the other player chooses its pure strategies with equal probabilities. In this case, there exists a class of deterministic policies that forces equality of the mean
payoffs per move of the players. This class of policies is the class of generalized tit-for-tat policies. When a representative player has a generalized tit-for-tat policy, in the limit of the iterated game, the payoffs of both representative players are equal. If a player tries to increase its payoff by changing its strategy and the other player plays tit-for-tat, the  change in the strategy can increase or decrease the payoffs, but  the payoffs per move of both players remain equal. Generalized tit-for-tat  
or imitation strategies force equalitarian payoffs per move. 
In dilemmatic games, the generalized tit-for-tat policy together with the condition of not being the first to defect, leads to the highest possible mean payoffs per move for the players.

\bigskip
{\bf Acknowledgments:} 
This   work has been partially supported by the  POCTI  Project P/FIS/13161/1998, and by Funda\c c\~ao para a 
Ci\^encia e a Tecnologia, under a plurianual funding grant.

\bigskip

\noindent{\bf References}
\bigskip

\noindent [1] J. von Neumann and O. Morgenstern, Theory of Games and Economic Behavior, Princeton University Press, 1944.
\smallskip

\noindent [2] J. Maynard Smith, Evolution and the Theory of Games, Cambridge University Press, 1982.
\smallskip

\noindent [3]  R. Axelrod, The evolution of cooperation, Basic Books, Harper Collins Pub., 1984.
\smallskip

\noindent [4] D. Fudenberg and J. Tirole, Game Theory, The MIT Press, Cambridge, Massachusetts, 1993.
\smallskip

\noindent [5] H. W. Kuhn (ed.), Classics in Game Theory, Princeton Uni. Press, 1997.
\smallskip

\noindent [6] F. Vega-Redondo, Evolution, Games and Economic Behaviour, Oxford Uni. Press, 1996.
\smallskip

\noindent [7] D. A. Meyer, Quantum strategies,
{\it Phys. Rev. Lett.}, {\bf 82} (1999) 1052-1055.
\smallskip

\noindent [8] C. A. Holt and A. E. Roth, The Nash equilibrium: A perspective, {\it Proc. Natl. Acad. Sci. USA}, {\bf 101} (2004) 3999-4002.
\smallskip

\noindent [9] P. D. Taylor and L. B. Jonker, Evolutionary stable strategies and game dynamics, {\it Math. Biosciences}, {\bf 40} (1978) 145-156.
\smallskip

\noindent [10] J. Hofbauer and K. Sigmund, Evolutionary Games and Population Dynamics, Cambridge Uni. Press, 1998.
\smallskip

\noindent [11] H. W. Kuhn , Lectures on Game Theory, Princeton Uni. Press, 2003.
\smallskip

\noindent [12] Mesterton-Gibbons, M., An introduction to Game-Theorethic Modelling, American Mathematical Society, Providence, RI, 2000.
\smallskip

\noindent [13] J. Nash, Equilibrium points in n-person games, 
{\it Proc. Natl. Acad. Sci. USA}, {\bf 36} (1950) 48-49.
\smallskip

\noindent [14] S. Smale, The prisoner's dilemma and dynamical systems associated to non-cooperative games, {\it Econometrica}, {\bf 48} (1980) 1617-1634.
\smallskip

\noindent [15] R. Tolman, The Principles of Statistical Mechanics, Dover, New York, 1979.
\smallskip

\noindent [16] V. I. Arnold and A. Avez, Problèmes Ergodiques de la Mécanique Classique, Gauthier-Villars, Paris, 1967.
\smallskip

\noindent [17] J. Maynard Smith and G. Price, The logic of animal conflicts, {\it Nature}, {\bf 246} (1973) 15-18.
\smallskip

\end{document}